\documentstyle[12pt]{article}
\newcommand{\be}{\begin{equation}}
\newcommand{\ee}{\end{equation}}
\newcommand{\bea}{\begin{eqnarray}}
\newcommand{\eea}{\end{eqnarray}}

\title{
On Low-Energy Effective Action in ${\cal N}=2$ Super Yang-Mills
Theories on Non-Abelian Background }

\author{A.T. Banin$^{1}$, I.L. Buchbinder$^{2}$\footnote{joseph@tspu.edu.ru}
, N.G. Pletnev$^{1}$\footnote{pletnev@math.nsc.ru}}

\date{{\it
${^{1}}$Institute of Mathematics, \\
Novosibirsk, 630090, Russia,\\
\vspace{0.7cm}
${^{2}}$Department of Theoretical
Physics, \\ Tomsk State Pedagogical University, \\ Tomsk, 634041,
Russia } }

\begin{document}

\begin{titlepage}
\maketitle
\begin{abstract}
We compute the non-holomorphic corrections to low-energy effective
action (higher derivative terms) in ${\cal N}=2$, $SU(2)$ SYM
theory coupled to hypermultiplets on a non-abelian background for
a class of gauge fixing conditions. A general procedure for
calculating the gauge parameters depending contributions to
one-loop superfield effective action is developed. The one-loop
non-holomorphic effective potential is exactly found in terms of
the Euler dilogarithm function for specific choice of gauge
parameters.

\end{abstract}
\thispagestyle{empty}
\end{titlepage}

\section{Introduction}

Low-energy effective action of ${\cal N}=2$ supersymmetric
Yang-Mills theories is defined, in purely gauge superfield sector,
by two effective potentials. The leading correction is given by
holomorphic potential ${\cal F}({\cal W})$ and the next-to-leading
correction is written in terms of non-holomorphic potential ${\cal
H}({\cal W}, \bar {\cal W})$ where ${\cal W}$ and $\bar {\cal W}$
are ${\cal N}=2$ superfield strengths (see e.g. the review
\cite{bko}).

${\cal N}=2$ supersymmetry strongly restricts the form of
holomorphic potential what was demonstrated by Seiberg and Witten
for $SU(2)$ SYM model in Coulomb branch of inequivalent vacua in
which the low energy theory has unbroken $U(1)$ gauge factors \cite{sw}.
Extension of this result for various gauge groups and coupling to matter was
given in ref. \cite{h} (see also the review \cite{hoph}).
General form of holomorphic potential for arbitrary ${\cal N}=2$
model is now well established.

Computation of non-holomorphic potential is more delicate and a
general form of ${\cal H}({\cal W}, \bar{\cal W})$ is still
unknown although some contributions to ${\cal H}$ were obtained
for special cases. In ${\cal N}=2$ superconformal invariant models
and ${\cal N}=4$ SYM theory the non-holomorphic potential has been
found in Coulomb phase \cite{wigr} -- \cite{lowe}. Here all beta
functions vanish and the evolutions under the renormalization group is
trivial. This effective potential is turned
out to be exact solution for ${\cal
N}=4$ SYM theory, its explicit form is given only by one-loop
contribution, any higher-loop or instanton corrections are absent
\cite{dise}, \cite{lowe} -- \cite{bupe}. However all above results
correspond to abelian background ${\cal W}$ and $\bar{\cal W}$ for
the theory, living  on a point of general position of the moduli
space, where one has the symmetry-breaking pattern:
$SU(N)\rightarrow U(1)^{N-1}$ and all physical quantities vary
smoothly over the moduli spaces. The moduli space becomes an
orbifold, so that it is flat "almost" everywhere else and has infinite
curvature at the origin. The singularities of the
moduli space are associated with the presence of new massless
particles in the spectrum \cite{ark}. Besides, there exists a curve of
marginal stability where otherwise stable BPS states become
degenerate and can decay into a very few strong-coupling states
\cite{bill}.  In addition, all such points have an enhanced
non-abelian symmetry which forms some non-abelian background. In
this region, unstability of the low-energy approximation is
expected to be broken down when the derivatives of scalar fields and
$U(1)$-field strength become large. In
particular, an analysis of such regions is important for
understanding a quantum corrected form of BPS solution in strong
coupling region \cite{chru}.  As to non-abelian background, the
non-holomorphic potential was found only for very special cases in
refs. \cite{wigr} -- \cite{grro}.

One of the basic approaches to evaluating the effective action is
a derivative expansion. This approach allows to get the effective
action in form of a series in derivatives of its functional
arguments. Within ${\cal N}=1$ supersymmetric derivative
expansion, the leading contributions to effective action are
formed by so called K\"{a}hlerian and chiral superfield effective
potentials. The K\"{a}hlerian effective potential has a structure
analogous to conventional effective potential, its form has been
recently investigated for various ${\cal N}=1$ supersymmetric
models. Supercovarint derivatives depending corrections to
K\"{a}hlerian effective potential can be found using the methods
developed in refs. \cite{bkyc} -- \cite{bapl}. We point out that
the K\"{a}hlerian effective potential naturally arises in ${\cal
N}=2$ SYM models if ones formulate these models in terms of ${\cal
N}=1$ superfields \cite{gat} and, as a result, it allows to construct the
potentials ${\cal F}({\cal W})$ and ${\cal H}({\cal W}, \bar{\cal
W})$ on its ground.

The powerful approach to evaluating the effective action in ${\cal
N}=2$ SUSY models can be developed within harmonic superspace \cite{gaiv}
since this superspace provides a formulation of ${\cal N}=2$
supersymmetric theories in terms of unconstrained ${\cal N}=2$
superfields and, therefore, preserves a manifest off-shell ${\cal N}=2$
supersymmetry. Structure of effective action of ${\cal N}=2$ and ${\cal
N}=4$ SYM models in harmonic superspace has been studied in refs
 \cite{buk, bupe, bbbi, kuar}.

Another line of current study of the effective action in extended SUSY
theories is associated with realization of these theories on the
world volume of branes. Such a realization provides a dual
description of low-energy field dynamics in terms of D-brane
theory. Webs of intersecting branes as a tool for studying the
gauge theories with reduced number of supersymmetries have been
introduced in ref. \cite{haw}. The fivebrane construction has been
successfully applied to the computation of holomorphic (or rather
BPS) quantities of the four dimensional supersymmetric gauge
theory (see refs. \cite{bho}, \cite{hlw}). The fivebrane
configurations corresponding to these ${\cal N} = 1$
supersymmetric gauge theories encode the information about the
${\cal N} = 1$ moduli spaces of vacua. The non-holomorphic
quantities such as higher derivative terms in ${\cal N}=2$
theories and the K\"ahlerian potential of ${\cal N}=1$
supersymmetric gauge theories have a special interest since they
are not protected by supersymmetry.  It was shown that the
K\"ahlerian potential on the Coulomb branch of ${\cal N}=2$
theories is correctly reproduced from the classical dynamics of
M-theory fivebrane. As to the non-holomorphic contributions to
low-energy effective action, such as the higher derivative terms,
a correspondence between string/brane approach and
four-dimensional ${\cal N}=2$ supersymmetric Yang-Mills theories
beyond two-derivative level, is not completely established (see
e.g. ref. \cite{bho, hlw, koch}). Moreover, in an arbitrary ${\cal
N}=2$ SYM model coupled to hypermultiplets, a general solution to
the function ${\cal H}$ and its derivative depending corrections
in the points of moduli space corresponding to non-abelian
background is still far to be found.

In this paper we discuss some aspects of structure of
non-holomorphic effective potential for nonabelian background
in order to pay attention on a problem of its gauge dependence. We show
that for unbroken $SU(2)$ gauge group the one-loop non-holomorphic
potential can be exactly calculated for a wide class of gauge
fixing conditions. It is generally known that the contributions to
the effective action, that contain factors of the classical equations of
motion, aren't uniquely defined. They are often ignored.  An example of this
ambiguity is a dependence of the effective action on the
choice of gauge conditions in a gauge theory. This fact is related
to the parameterization non-invariance of the conventional
effective action (see e.g. ref. \cite{gosp}) and leads to a number
of different effective actions corresponding to the one classical
action.

We present an extended supersymmetrical $R_{\xi}$-gauge for SYM models
within background field method. The choice of a gauge fixing term in
spontaneous broken non-abelian gauge theories has a basic technical importance.
It is known that the use of the $R_{\xi}$-gauge became a major step in
the proof that Yang-Mills models are unitary, on-shell gauge
independent and renormalizable quantum field theories.
One of our motivation is to provide a useful "laboratory" for studying
a full structure of low-energy EA in hypermultiplet model coupled
non-abelian ${\cal N}=2$ vector multiplet.

The structure of the paper is as follows:\\
In the second section we introduce general notations and remind the known
procedure of reduction ${\cal N}=2$ superfields and action to ${\cal N}=1$ superspace.
Section three presents the background field quantization method for the
model under consideration.  The extension of $R_{\xi}$-gauge fixing for
the model is also introduced in this section.  In the forth section we
study the gauge-dependence of ${\cal N}=1$ superfield the K\"ahler
potential and consider the problem of reconstruction ${\cal H}$.
In short summary we discuss the results obtained.

\section{${\cal N}=2$ SYM Theory in ${\cal N}=1$ Superspace}
A most simple and well developed description of four-dimensional
supersymmetric field theories is formulation in terms of ${\cal
N}=1$ superspace. Although the ${\cal N}=2$ supersymmetric models
can be constructed in harmonic superspace \cite{gaiv} preserving
manifest ${\cal N}=2$ supersymmetry, the ${\cal N}=1$ formulation
is still very useful and fruitful for study of the various quantum
aspects in the ${\cal N}=2$ supersymmetric models.

From point of view of ${\cal N}=1$ supersymmetry a field content
of pure ${\cal N}=2$ SYM model is given by vector multiplet
superfiled $V$ and chiral superfield ${\Phi}$ and a field content
of hypermultiplet is given by two chiral superfields $Q_{+},
\bar{Q}_{-}$. This allows to write an action $S$ of the ${\cal
N}=2$ SYM model coupled to hypermultiplet matter in ${\cal N}=1$
superspace as follows \bea
S&=&S_{\rm SYM} + S_{\rm Hyper}\\
S_{\rm SYM}&=&{1 \over T(R)g^2}{\rm tr}[\int d^6z\, {1 \over
2}W^{\alpha}W_{\alpha} + \int
d^8z\, \bar{\Phi}{\rm e}^{V}\Phi {\rm e}^{-V}], \label{clas}\\
S_{\rm Hyper}&=& \int d^8z \,(\bar{Q}_{+}{\rm e}^{V}Q_{+}+Q_{-}
{\rm e}^{-V}\bar{Q}_{-}) + i\int d^6z \, Q_{-}\Phi Q_{+} + i\int
d^6\, \bar{z}\bar{Q}_{+}\bar{\Phi}\bar{Q}_{-},\label{clhp}
\eea
where the superfields $V=V^{A}T^{A}$ and $\Phi=\Phi^{A}T^{A}$ form
the ${\cal N}=2$ gauge multiplet with component fields $(A_{\mu},
\lambda_{\pm}, \phi)$ belonging to the adjoint representation of gauge
group $G$ and the superfields $Q_{\pm}$ form the hypermultiplet
with component fields $(\psi_{+},H_{\pm}, \psi_{-})$ belonging to
some representation ${\cal R}$ of $G$. We use the conventions of
ref. \cite{ggrs}. It should be noted that the used gauge coupling
constant $g$ is $\sqrt{\,2}$ times the usual $g$. The $T^{A}$ are
the generators of a gauge group with $[T^{A},T^{B}]=if^{ABC}T^{C}$.
These generators satisfy the normalizing conditions
${\rm tr}(T^{A}T^{B})=T(R)\delta^{AB}$,
$(T^{A})_{ij}(T^{A})_{jk}=C(R)\delta_{ik}$ and
$f^{ACD}f^{BCD}=C_{2}(G)\delta^{AB}$. The term $Q_{-}\Phi Q_{+}$
in the Lagrangian (\ref{clhp}) means $Q_{-i}(T^{A})_{ij}Q_{+j}\Phi^{A}$.

The classical actions $S_{\rm SYM}$ and $S_{\rm Hyper}$ are gauge
invariant and  manifestly ${\cal N}=1$ supersymmetric by
the construction. However the full action $S$ is also
invariant under the hidden ${\cal N}=2$ supersymmetry transformations,
which can be written in terms of covariant chiral superfields
$\Phi_{c}={\rm e}^{\bar\Omega}\Phi{\rm e}^{-\bar\Omega}$,
$Q_{+ c}= {\rm e}^{\bar\Omega}Q_{+}$ etc.
\bea
&\delta\Phi_{c}=\epsilon^{\alpha}W_{\alpha},\quad
\delta\bar{\Phi}_{c}=\bar{\epsilon}^{\dot{\alpha}}\bar{W}_{\dot{\alpha}},&
\nonumber\\
& \delta W_{\alpha}=-\epsilon_{\alpha}\bar{\nabla}^{2}\bar{\Phi}_{c}+
i\epsilon^{\dot{\alpha}}\nabla_{\alpha\dot{\alpha}}\Phi_{c}, \quad
 \delta\bar{W}_{\dot{\alpha}}=-\bar{\epsilon}_{\dot{\alpha}}\nabla^{2}\Phi_{c}+
i\epsilon^{\alpha}\nabla_{\alpha\dot{\alpha}}\bar{\Phi}_{c},&\label{trans}
\eea

\bea
&\delta\bar{Q}_{+\,c}=\bar{Q}_{+\,c}(\Delta_{1} \Omega)- \nabla^{2}
(Q_{-\,c}\chi),
\quad \delta\bar{Q}_{-\,c}=-(\Delta_{1} \Omega)\bar{Q}_{-\,c}
+\nabla^{2}(\chi Q_{+\,c}),& \nonumber\\
&\delta Q_{+\,c}=- (\Delta_{2} \Omega)Q_{+\,c} +\bar{\nabla}^{2}(\chi
\bar{Q}_{-\,c}), \quad \delta Q_{-\,c}=Q_{-\,c}(\Delta_{2} \Omega)-
\bar{\nabla}^{2}(\bar{Q}_{+\,c}\chi),& \nonumber\\
&\Delta_{1} \Omega = {\rm e}^{-\Omega}\delta {\rm e}^{\Omega}=
i\chi\Phi_{c},
\quad \Delta_{2} \Omega = {\rm e}^{\bar\Omega}\delta
{\rm e}^{-\bar\Omega}= i\bar{\Phi}_{c}\chi,& \nonumber\\
&\chi= \lambda(\theta)+ \bar{\lambda}(\bar\theta),&\label{transa}
\eea
Here $\Omega$ is a complex superfield determining the gauge superfield
$V$ in the form ${\rm e}^{V}={\rm e}^{\Omega}{\rm e}^{\bar{\Omega}}$,
$\lambda$ and $\bar\lambda$ are chiral and antichiral space-time
independent superfield parameters with the expansion
$\lambda=\gamma + {1\over 2}\theta^{\alpha}\epsilon_{\alpha}+
\theta^{2}(\beta_{1}+i\beta_{2})$,
where the $\beta_{1},\,\beta_{2}$ parameterize the $SU(2)/U(1)$ group,
$\epsilon_{\alpha}$ are the anticommuting parameters presenting in
the eqs (\ref{trans}) and $\gamma$ parameterizes the central charge
transformations. Hypermultyplet action
and corresponding ${\cal N}=2$ supersymmetry transformations in terms
of ${\cal N} =1$ superspace were considered in refs.
\cite{ggrs} and \cite{galp}. Invariance of the actions $S_{SYM}$
and $S_{Hyper}$ under the transformations (\ref{trans}, \ref{transa}) can be checked
straightforwardly. One points out also that both ${\cal N}=2$ super
Yang-Mills model and hypermultiplet model are the superconformal
invariants \cite{bkt}. Further we will use only the covariant chiral
superfields and subscript $c$ will be omitted.

Low-energy effective action of the model under consideration is
described by holomorphic scale dependent effective potential
${\cal F({\cal W})}$ and non-holomorphic scale independent real
effective potential ${\cal H}({\cal W}, \bar{\cal W})$ where
${\cal W}$ is ${\cal N}=2$ superfield strength. The corresponding
contributions to effective action can be expressed in terms of
${\cal N}=1$ superfields. The holomorphic part $\Gamma_{F}$ of
low-energy effective action is written in ${\cal N}=1$ form as
follows \cite{gat}
\be\label{holp}
\Gamma_{F}= \int d^{4}x d^{2}\theta\,
{1\over 2}{\cal F}_{AB}(\Phi)W^{A\alpha} W^{B}_{\alpha} + \int
d^{4}x d^{4}\theta\,{\cal F}_{A}(\Phi)\bar{\Phi}^{A}+ h.c.
\ee
We use the standard notation
${\cal F}_{A}={\partial\over\partial
\Phi^{A}}{\cal F}$, ${\cal F}_{AB}={\partial\over\partial
\Phi^{A}}{\partial\over\partial \Phi^{B}}{\cal F}$, ${\cal
H}_{A}={\partial\over\partial \Phi^{A}}{\cal H}$, ${\cal
H}_{A\bar{B}}={\partial\over\partial
\Phi^{A}}{\partial\over\partial \bar{\Phi}^{B}}{\cal H}$ etc.
The non-holomorphic contribution $\Gamma_{H}$ can be given in
${\cal N}=1$ form using the metric, connection and curvature of
natural K\"ahler geometry since the ${\cal H}$ is associated with
a K\"ahler potential on a complex manifold defined modulo the
real part of a holomorphic function
\bea
\Gamma_{H}&=&\int d^{4}x d^{4}\theta\,
(g_{A\bar{B}}[-{1\over 2}\nabla^{\alpha\dot{\alpha}}\Phi^{A}
\nabla_{\alpha\dot{\alpha}}\bar{\Phi}^{B}+
i\bar{W}^{B\dot{\alpha}}(\nabla^{\alpha}_{\dot{\alpha}}W^{A}_{\alpha}+
\Gamma^{A}_{CD}\nabla^{\alpha}_{\dot{\alpha}}\Phi^{C}W^{D}_{\alpha})-\nonumber\\
&-&(f^{ACD}\bar{W}^{B\dot{\alpha}}\Phi^{C}\bar{\nabla}_{\dot{\alpha}}\bar{\Phi}^{D}+
f^{BCD}W^{A\alpha}\bar{\Phi}^{C}\nabla_{\alpha}\Phi^{D})+
(\nabla^{2}\Phi^{B}+{1\over
2}\Gamma^{\bar{B}}_{\bar{C}\bar{D}}\bar{W}^{C\dot{\alpha}}
\bar{W}^{D}_{\dot{\alpha}})\times\nonumber\\
&\times&(\bar{\nabla}^{2}\bar{\Phi}^{A}+{1\over 2}\Gamma^{A}_{EF}
W^{E\alpha}W_{\alpha}^{F})]+ {1\over
4}R_{A\bar{B}C\bar{D}}(W^{A\alpha}W^{C}_{\alpha}\bar{W}^{B\dot{\alpha}}
\bar{W}^{D}_{\dot{\alpha}})+\nonumber\\
&+& i\mu^{A}({1\over 2}\nabla^{\alpha}W^{A}_{\alpha}+
f^{ABC}\Phi^{B}\bar{\Phi}^{C})),\label{full} \eea where the last
term in (\ref{full}) written in terms (see ref. \cite{wigr}) of
the moment map (or Killing potential)
$i\mu^{A}(\Phi,\bar{\Phi})=f^{ABC}{\cal H}_{C}\Phi^{B}$ and
$g_{A\bar{B}}={\cal H}_{A\bar{B}}$,
$\Gamma^{A}_{BC}=g^{A\bar{D}}{\cal H}_{BC\bar{D}}$,
$R_{A\bar{B}C\bar{D}}={\cal
H}_{AC\bar{B}\bar{D}}-g_{E\bar{F}}\Gamma^{E}_{AC}\Gamma^{\bar{F}}_{\bar{B}\bar{D}}$.
The momentum map explains the nature of the auxiliary fields and
it is used to write down the scalar potential of ${\cal N}=2$
theories. It should be noted that on ${\cal N}=1$ language the
representation (\ref{full}) for the non-holomorphic potential
${\cal H}({\cal W},\bar{\cal W})$ is essentially as expansion over
derivatives $W,\, \Phi$ and an arbitrary number of external $\Phi$
superfields. Expression (\ref{full}) can be also found directly
from eq. (\ref{expa}). Coefficients of the expansion (\ref{full})
containing derivatives ${\cal H}$ can be written by means of
natural for K\"ahler geometry unit vectors
$e^{A}=\Phi^{A}/\sqrt{\Phi^{2}}$,
$e^{\bar{A}}=\bar{\Phi}^{A}/\sqrt{\bar{\Phi}^{2}}$, projectors
$\Pi^{AB}=\delta^{AB} - e^{A}e^{B}$,
$\Pi^{\bar{A}\bar{B}}=\delta^{\bar{A}\bar{B}} -
e^{\bar{A}}e^{\bar{B}}$ and their derivatives. Being expressed in
terms of component fields, the contribution to effective action
$\Gamma_{H}$ contains at most four space-time derivatives.
Obtaining the non-holomorphic contribution to the effective action
in a form of an integral over ${\cal N}=1$ superspace is based on
decomposition of the non-abelian superfield strengths ${\cal W},
\bar{\cal W}$ in terms of ${\cal N}=1$ superfields, which is given
in the {\it Appendix}.

In this paper we analyze a general form of the
one-loop functionals $\Gamma_{F}$ and $\Gamma_{H}$ in the model
under consideration using functional methods in ${\cal N}=1$ superspace and
revise the contributions to the effective action which determine a
functional dependence of ${\cal F}$ and ${\cal H}$ on the ${\cal N}=2$
vector multiplet.  Eqs (\ref{holp},\ref{full}) play a very important
role in such an approach since they ensure a bridge between ${\cal N}
=1$ and ${\cal N}=2$ descriptions and allow to restore manifestly
${\cal N}=2$ supersymmetric functionals on the base of their ${\cal
N}=1$ projections.

Calculations of the low-energy effective action are based on the
following reasoning:
We compute one-loop contributions to the effective
potential $K(\Phi, \bar\Phi)$ induced by both ${\cal N}=2$ vector
multiplet and hypermultiplet in a wide class of gauge-fixing
conditions. This effective potential depends on ${\cal N}=1$ chiral
superfields which are a part of the ${\cal N}=2$ vector multiplet. The
functionals (\ref{holp},\ref{full}) also contain terms depending
only on $\Phi, \bar\Phi$ and this allows to restore such terms on
the base of the given effective potential $K(\Phi, \bar\Phi)$.
It is known that this effective potential can not be written in the
form $\bar\Phi {\cal F}'(\Phi)$, which saturates the R-anomaly.
The additional scale independent terms in the effective potential $K$
originate from a real function ${\cal H}$. This function ${\cal H}$
can be determined from comparison of the last term in (\ref{full}) and
effective potential $K$. The other terms in (\ref{full}) arise from a
momentum expansion in ${\cal N}=1$ superspace and related by extra hidden
${\cal N}=1$ supersymmetry to one another. Therefore, they
can be exactly found on the base of the terms depending only on $\Phi, \bar\Phi$.

\section{Background Field Quantization}

\subsection{Quantum-background splitting}
To construct the effective action in the ${\cal N}=1$ SYM theory
with matter multiplets we use the background field method which is
a powerful and convenient tool for studying the structure of a
quantum gauge theory (see refs. \cite{ggrs, idea}). This method
begins with the so-called background-quantum splitting of the
initial gauge and matter superfields into two parts --- into
background superfields and the quantum superfields, according to
the following transformation ${\rm e}^{V} \to {\rm e}^{\Omega}{\rm
e}^{V}{\rm e}^{\bar\Omega}$ and $\Phi \to \Phi+\phi$ in the
actions (\ref{clas}, \ref{clhp}). As a result, these actions will be
written as the functionals of the background superfields
${\Omega}, {\bar\Omega}, {\Phi}, {\bar\Phi}$ and quantum ones $V,
{\phi}, {\bar\phi}$.

To quantize the
theory we impose the gauge-fixing conditions only on the quantum
fields, introduce the corresponding ghosts and consider the
background fields as the functional arguments of the effective action.
Using the proper gauge fixing functions
one can construct the effective action which will be invariant
under the initial classical gauge transformations. Due to this property
the effective action depends only on background strengths
$W_{\alpha}$ and $\bar{W}_{\dot\alpha}$, covariantly-chiral superfields
$\Phi$ and $\bar{\Phi}$ and their covariant derivatives.

The gauge transformations of the quantum superfields $\phi$ and $V$
are written as follows
\bea
& \phi'={\rm e}^{i\Lambda}(\Phi+\phi){\rm e}^{-i\Lambda}-\Phi, \quad
\bar{\phi}'={\rm e}^{i\bar{\Lambda}}(\bar{\Phi}+\bar{\phi}){\rm
e}^{-i\bar{\Lambda}}-\bar{\Phi},& \nonumber\\
&\delta V=i(\bar{\Lambda}-\Lambda)-{i\over
2}[V,\bar{\Lambda}+\Lambda]+O(V^{2}),\label{qtr}&
\eea
Namely these transformations must be fixed by proper gauge conditions
imposed on the quantum superfields.

\subsection{Gauge-fixing procedure}
The basic step of the background field method is use
of the gauge fixing conditions which are covariant under
the background gauge transformations.
We choose the proper gauge-fixing conditions for the quantum
superfields  $V$ and ${\phi}$ in the form
\be
\label{fix}
\bar{F}^{A}=\nabla^2V^{A} +i\lambda {1\over
\Box_{+}}\nabla^{2}\phi^{B}\bar{\Phi}^{C}f^{ABC},
\quad
F^{A}=\bar{\nabla}^2V^{A} -i\bar{\lambda} {1\over
\Box_{-}}\bar{\nabla}^2\bar{\phi}^{B}\Phi^{C}f^{ABC},
\ee
where $\lambda, \bar{\lambda}$ are the arbitrary numerical parameters and
standard notations $\Box_{\pm}$ for Laplace-like operators in the
superspace are used. In space of chiral and antichiral superfields
these operators act accordingly
\be
\label{box}
\nabla^{2}\bar{\nabla}^{2}= \Box_{+}=
\Box-i\bar{W}^{\dot{\alpha}}\bar{\nabla}_{\dot{\alpha}}-
{i\over 2}(\bar{\nabla}\bar{W}),
\quad
\bar{\nabla}^{2}\nabla^{2}= \Box_{-}=
\Box-i W^{\alpha}\nabla_{\alpha}-{i\over 2}(\nabla W).
\ee
It is evident that the gauge fixing functions (\ref{fix}) are covariant under
background superfield transformations.
The gauge fixing functions (\ref{fix}) can be considered as a superfield form
of so called $R_{\xi}$-gauges which are used often in spontaneously
broken gauge theories. Extension of $R_{\xi}$-gauge fixing conditions to
${\cal N}=1$ superfield theories has been given in ref. \cite{ovw}.

Gauge fixing action corresponding to the functions (\ref{fix}) is constructed
in the standard form
\be\label{fixg}
S_{\rm GF}=-{1\over \alpha g^{2}}\int d^8z\,(F^{A}\bar{F}^{A}+b^{A}\bar{b}^{A})
\ee
and depends on extra parameter ${\alpha}$. Invariance of this action
under the background gauge transformations is evident

Action of the Faddeev-Popov ghosts $S_{FP}$ for the gauge fixing
functions (\ref{fix}) has the form
\be
S_{\rm FP}= {\rm tr}\int d^8z \,\left(
(\bar{c}'c-c'\bar{c})-
\left(
c' [\Phi,{\lambda\over \Box_{+}} [\bar{c},\bar{\Phi} ] ]+
\bar{c}' [{\bar\lambda\over \Box_{-}} [c,\Phi ],\bar{\Phi} ]
\right)\right).\label{acfp}
\ee
The theory under consideration demands, besides Faddeev-Popov ghosts,
the extra, so called Nielsen-Kallosh ghosts $b$ and $\bar{b}$. The
full ghost action be
\be\label{sgh}
S_{\rm GH} = S_{\rm FP} + {\rm tr}\int d^8z\, b\bar{b}
\ee
The actions (\ref{fixg}, \ref{sgh}) can be rewritten in more convenient form
if we introduce the following $\Phi$-dependent denotations:
\be\label{xab}
X^{AB}= f^{ACB}\Phi^{C},\quad \bar{X}^{AB}=f^{ACB}\bar{\Phi}^{C}.
\ee
Using these denotations, integrating
by part and dropping the irrelevant for one-loop calculation
terms ones get the expression for the gauge fixing action
\bea
S_{\rm GF}&=& -{1\over \alpha g^{2}} \int d^{4}\theta\, (
\nabla^{2}V^{A}\bar{\nabla}^{2}V^{A}+i\lambda V^{A}\bar{X}^{AB}\phi^{B}-
i\bar{\lambda}V^{A}X^{AB}\bar{\phi}^{B}+\nonumber\\
&+&\lambda\bar{\lambda}\nabla^{2}{1\over \Box_{-}}\,\phi^{B}\bar{X}^{BA}
\bar{\nabla}^{2}{1\over \Box_{+}}\,\bar{\phi}^{E}X^{AE}\label{gfix}
)
\eea
We point out, because the parameter $\Lambda$ of the quantum
field transformations (\ref{qtr}) is chiral the ghosts $c$,
$c'$ and $b$ are covariant chiral superfields.
$\bar{\nabla}_{\dot{\alpha}}c=\bar{\nabla}_{\dot{\alpha}}c'=
\bar{\nabla}_{\dot{\alpha}}b=0$.

The quadratic part of the full ghost action (\ref{sgh}) which is relevant for
one-loop calculations can be also given in terms of the denotions (\ref{xab})
\be\label{fpac} S_{ghost}=\int d^8z\,
(\bar{c}^{'A}c^{A}-c^{'A}\bar{c}^{A}+
\bar{c}^{B}{\lambda\over\Box_{-}}\bar{X}^{BE}X^{EA}c^{'A}+
\bar{c}^{'B}{\bar{\lambda}\over\Box_{-}}\bar{X}^{BE}X^{EA}c^{A}+b^{A}\bar{b}^{A}).
\ee

To carry out the loop calculations, we expand the total action
$S=S_{\rm SYM}+S_{\rm Hyper}+S_{\rm GH}+S_{\rm GF}$ in power series in quantum
fields. Only quadratic terms in this expansion are relevant in
one-loop approximation. The corresponding quadratic part of total
action can be written as follows
\be
S_{2}=S_{gauge}+S_{chiral}+S_{mix}+S_{ghost}.
\ee
The contributions from the quantum field $V$ is given by
\be\label{gaac}
 S_{gauge}=-{1\over 2g^{2}T(R)}{\rm tr}\int
d^{4}\theta\,
V^{A}\left(\Box-iW^{\alpha}\nabla_{\alpha}-i\bar{W}^{\dot{\alpha}}\bar{\nabla}_{\dot{\alpha}}-
(1-\alpha^{-1})\{\nabla^{2},\bar{\nabla}^{2}\}- M \right)^{AB}V^{B},
\ee
where $\Box={1\over 2}\nabla^{\alpha\dot{\alpha}}\nabla_{\alpha\dot{\alpha}}$ is the
background covariant d'Alambertian and $W_{\alpha}, \bar{W}_{\dot{\alpha}}$ are
the background field strengths. The mass matrix
$M^{BA}= -{1\over 2}\bar{X}^{(BE}X^{EA)}$ in the action $S_{gauge}$ arises from term
${1\over 2}\bar\Phi[V,[V,\Phi]]$ in the action $S_{\rm SYM}$ (see (\ref{clas})).
The action $S_{mix}$ contains terms mixing the quantum gauge and the
chiral superfields:
\be
S_{mix}=\int d^{4}\theta\, (\bar{\phi}[V, \Phi]+[\bar{\Phi}, V]\phi).
\ee

The action $S_{chiral}$ is quadratic in the quantum chiral superfield $\phi$ and
$Q$
\be\label{acch}
S_{chiral} = {\rm tr}(\int d^{4}\theta\,\bar{\phi}\phi +\int
d^{4}\theta\,(\bar{Q}_{+}Q_{+}+Q_{-}\bar{Q}_{-})+ i\int
d^{2}\theta\,Q_{-}\Phi Q_{+} +i\int
d^{2}\bar{\theta}\,\bar{Q}_{+}\bar{\Phi}\bar{Q}_{-}).
\ee
All one-loop contributions to effective action are given in terms of
the functional trace ${\rm Tr}\ln (\hat{H})$, where the operator $\hat{H}$
is the matrix of the second variational derivatives of the action $S_{2}$
in all quantum fields. The one-loop effective action in the
model under consideration reads
\be
\label{ggam} \Gamma [V,\Phi]={i\over 2}{\rm Tr}\ln
\hat{H}_{\rm SYM}+i{\rm Tr}\ln \hat{H}_{\rm Hyper} - {i\over 2}{\rm Tr}\ln
\hat{H}_{ghost},
\ee
with
\be\label{hsym} \hat{H}_{\rm SYM} =\pmatrix{
{(O_{V})}^{BA}& i\bar{\gamma}X^{BA}\nabla^2
&-i\gamma\bar{X}^{BA}\bar{\nabla}^2 \cr
i\gamma\bar{\nabla}^2\bar{X}^{BA}&{(1+\bar{R})}^{BA}\bar{\nabla}^2 \nabla^2 & 0 \cr
-i\bar{\gamma}\nabla^2 X^{BA} & 0 &{(1+R)}^{BA}\nabla^2 \bar{\nabla}^2}
\ee

\be\label{hhyp}
\hat{H}_{\rm Hyper}=\pmatrix{\delta_{i}^{j}\bar{\nabla}^2 \nabla^2 & i\Phi_{i}^{j}\bar{\nabla}^2\cr
i\bar{\Phi}_{i}^{j}\nabla^2 & \delta_{i}^{j}\nabla^2 \bar{\nabla}^2 }
\ee

\be\label{hgho}
\hat{H}_{ghost} = \pmatrix{
0 &\nabla^2 (1 + G)\bar{\nabla}^2 & 0& 0\cr
-\bar{\nabla}^2 (1 + G^{\rm T})\nabla^2 & 0 & 0 & 0\cr
0 & 0 & 0 &\nabla^2 (1 + \bar{G})\bar{\nabla}^2 \cr
0 & 0 & -\bar{\nabla}^2 (1 + \bar{G}^{\rm T})\nabla^2 & 0 },
\ee
where we use the following notation
$\bar{R}^{BA}={\lambda\bar{\lambda}\over \alpha}\bar{X}^{BE}X^{EA}{1\over
\Box_{-}}$, $R^{BA}={\lambda\bar{\lambda}\over
\alpha}X^{BE}\bar{X}^{EA}{1\over \Box_{+}}$,
$G^{AB}={\bar\lambda\over\Box_{-}}\bar{X}^{AE}X^{EB}$, $\bar{G}^{AB}={\lambda\over\Box_{+}}\bar{X}^{AE}X^{EB}$ and
$O_{V}= -\Box+iW^{\alpha}\nabla_{\alpha}+i\bar{W}^{\dot{\alpha}}\bar{\nabla}_{\dot{\alpha}}+
(1-\alpha^{-1})\{\nabla^{2},\bar{\nabla}^{2}\}+M$.
Constants $\gamma$ and $\bar{\gamma}$ are defined as
$\bar{\gamma}=(1-\bar{\lambda}/\alpha),\,\gamma=(1-\lambda/\alpha)$.
The operator $\hat{H}_{SYM}$ contains the contributions from
${\cal N}=1$ vector and chiral multiplets forming ${\cal N}=2$ gauge
multiplet. One can see, that choice $\lambda=\bar\lambda=\alpha$
greatly simplifies all calculation because it diagonilizes the matrix
$\hat{H}_{SYM}$ and decouples the contributions from ${\cal N}=1$
vector and chiral multiplets. However we will keep the gauge parameters
$\lambda$ and $\alpha$ arbitrary and investigate an dependence of
effective action on these parameters.

Since the EA is expressed in form of functional determinant of the differential
operator $\hat{H}$, its calculation can be carried out on the base of
Fock-Schwinger proper-time technique appropriately formulated in superspace
(see aspects of such a formulation in refs. \cite{arga, bapl, idea}).

Exact calculations of the functional traces defining the one-loop
effective action is possible only for very specific backgrounds when
eigenvalues and eigenfunctions of the operators under consideration are
known, that is rather exception then a rule. Further we will use a derivative
expansion of the effective action based on an symbol operator technique
adapted to ${\cal N}=1$ supersymmetric field models (see \cite{bapl} for details).
Our purpose is the calculations of the leading and subleading low-energy
contributions to the one-loop effective action.

\section{${\cal N}=1$ K\"ahler and Non-holomorphic ${\cal N}=2$
Potentials }
\subsection{${\cal N}=1$ K\"ahler potential}
In this section we study the form of the non-abelian low-energy
effective action $\Gamma=\int d^{8}z\, K$ and its gauge
dependence. We compute the one-loop contributions to K\"ahler
effective potential $K$ induced by both ${\cal N}=2$ vector
multiplet and hypermultiplet. It is known that in the non-abelian
case the K\"ahler potential cannot be written in the form $Im
(\bar{\Phi}{\cal F}'(\Phi))$ consistent with the rigid version of
special geometry (see e.g. refs. \cite{wigr, piwe}). The
additional terms originate from a real function ${\cal
H}_{0}({\cal W},\bar{\cal W})$ of the ${\cal N}=2$ YM superfield
strength ${\cal W}$. The results obtained in the present paper are
more general in compare with ones obtained in refs. \cite{wigr,
 grro, piwe} since we have used here the more general and
complicated gauges.

We study the one-loop effective action for $SU(2)$-gauge model
described by (\ref{clas}), (\ref{clhp}) and (\ref{fixg}) with
$R_{\xi}$-gauge fixing (\ref{fix}) and Faddeev-Popov
(\ref{fpac}) terms in the case when the gauge vector field $V$ is
purely quantum. We find K\"ahler potential in ${\cal N}=1$
superspace, and then, the holomorphic ${\cal F}$ and nonholomorphic
${\cal H}$ potentials in ${\cal N}=2$ superspace.
To calculate these potentials we consider the diagrams with external
$\Phi,\bar{\Phi}$ lines corresponding only to the constant field
background.  Such a choice of background superfields leads to a number
of technical simplifications due to the absence of the background gauge
field, which allows us replace all background covariant
derivatives by flat ones (i.e.  $\nabla\rightarrow D,
\bar{\nabla}\rightarrow \bar{D}$). This provides a possibility to use
the superspace projectors $P_{1} = {1 \over \Box}\bar{D}^2 D^2, \quad
P_{2} = {1 \over \Box}D^2 \bar{D}^2, \quad P_{T}= -{1 \over \Box}D
\bar{D}^2 D$ and $\Pi_{0}= P_{1}+P_{2}$ and simplify the evaluations
of the functional determinants (\ref{ggam}).

Eq (\ref{hgho}) allows us to write the ghost contribution as
follows
\be\label{trgh}
{\rm Tr}\ln (H_{\rm ghost})= {\rm Tr}\left(\ln(1+G)+
\ln (1+\bar{G})\right)\Pi_{0}.
\ee
Notation ${\rm Tr}(\cdots)$ means ${\rm tr}\int d^{8}z\,(\cdots)$ as
usual. Matrices $R$ and $G$ from the (\ref{hsym}) and (\ref{hgho}) are
expressed in terms of $X$ and $M$. Using the identities ${\rm tr}\ln
O=\ln \det O$ and $ \det O={1\over N!}\epsilon^{ab\ldots}
\epsilon_{cd\ldots}O^{c}_{a}O^{d}_{b}\ldots$ one can obtain
(\ref{trgh})
\be\label{figh}
{\rm Tr}\ln (H_{\rm ghost}) =
2\int d^{8}z\,\left(\ln \left( 1-{\lambda\over\Box}\bar{\Phi}\Phi\right) + \ln
\left(
1-{\bar{\lambda}\over\Box}\bar{\Phi}\Phi\right)\right)\Pi_{0},
\ee
where $\bar{\Phi}\Phi$ means the scalar product in isospin space.

The contribution of the hypermultiplet to the effective action for any
representation of gauge group is given by
\be\label{trhy}
{\rm Tr}\ln (H_{\rm Hyper})={1\over 2}{\rm Tr}\ln (1+
\pmatrix{
{1\over \Box}\Phi\bar{\Phi}P_{2}&0\cr
0&{1\over \Box}\bar{\Phi}\Phi P_{1}}
) ={1\over 2}{\rm Tr}\int d^{8}z\,\ln
\left(1+{\Phi\bar{\Phi}\over \Box} \right)_{i}^{j}\Pi_{0}
\ee
where the trace is taken over the representation of the hypermultiplet.
The eigenvalues of the matrix $(\Phi\bar{\Phi})^{i}_{j}$ containing in
the definition of hypermutiplet contribution ${\rm Tr}\ln H_{\rm
Hyper}$ in the fundamental representation are
$
((\bar{\Phi}\Phi)\pm \sqrt{((\bar{\Phi}\Phi)^2
-\Phi^2\bar{\Phi}^2)\mathstrut})/4.
$
For adjoint representation we have
\be
{\rm Tr}\ln (H_{\rm Hyper}^{adj})={1\over 2}{\rm Tr}\ln
\left(1-{X\bar{X}\over \Box} \right)_{A}^{B}\Pi_{0}=
\int d^{8}z\,\ln (1-{(\Phi\bar{\Phi})\over \Box})\Pi_{0}.
\ee

The other contributions in (\ref{ggam}) are given by
\bea
& &{\rm Tr}\ln H_{\rm SYM} = {\rm Tr}[\ln
(1-{M\over\Box})P_{\rm T} + \label{hfin}\\
& &+\ln \left(1-{1\over \Box}(\bar{\Phi}\Phi)(\lambda+\bar{\lambda})+
\lambda\bar{\lambda}\left((\bar{\Phi}\Phi)\over\Box \right)^2 \right)\Pi_{0} +\nonumber\\
& &+\ln (1-{1\over\Box}(\bar{\Phi}\Phi)(\lambda+\bar{\lambda})+
\lambda\bar{\lambda}\left((\bar{\Phi}\Phi)\over \Box \right)^2 -
{1 \over 2}{\lambda\bar{\lambda}\over \Box^2}((\bar{\Phi}\Phi)^2 -\Phi^2\bar{\Phi}^2) +\nonumber\\
& &+{\alpha\over \Box^2}((\bar{\Phi}\Phi)^2-\Phi^2\bar{\Phi}^2)
\left( {\lambda\bar{\lambda}\over -4\Box}
(\bar{\Phi}\Phi)+ {\lambda+\bar{\lambda}\over 2}-{\alpha\over 4}\right) )\Pi_{0}].\nonumber
\eea
The computations for the first term in (\ref{hfin}) lead to
\be\label{vect}
{\rm Tr}\ln (1-{M\over\Box})P_{\rm T}=\ln
\left(1-{\Phi\bar{\Phi}\over\Box }\right)P_{\rm T} + \ln \left(
1-{\bar{\Phi}\Phi\over \Box}+{((\bar{\Phi}\Phi)^2
-\Phi^2\bar{\Phi}^2)\over 4\Box^2} \right)P_{\rm T}
\ee

Taking into account the results above one can obtain the
K\"ahler potential in the model under consideration. For actual
computation we use the technique which was described in detail in ref.
\cite{bapl}. According to this technique in the case
under consideration it is sufficient to fulfil the following
replacements
\be
\Box\rightarrow -k^2,\quad \int d^{4}k \rightarrow
i\int_{0}^{\infty} {k^{2}dk^{2}\over (4\pi)^2},\quad
\Pi_{0}\rightarrow-{2\over k^2}, \quad P_{\rm T}\rightarrow{2\over
k^2},
\ee
and integrate over $k^2$. The momentum integral is divergent and needs
regularization. All cut-off dependence contributes only to
renormalization of the initial action. The typical integrals
are given by following expression
\be\label{integ}
\int_{0}^{\Lambda^2} dk^{2}\,\ln\left(1+{A\over k^{2}}\right)=-A\ln {A\over {\rm e }
\Lambda^2}
\ee
The final result is a sum of three terms:\\
1) The hypermultiplet contribution to effective action
\bea
K_{\rm Hyper}^{fund}&=&-{1 \over (4\pi)^2}\int dk^{2}\,\ln (1 + {(\bar{\Phi}\Phi)+
\sqrt{((\bar{\Phi}\Phi)^2 -\Phi^2\bar{\Phi}^2)\mathstrut}\over
4k^2})-\nonumber\\
&-&{1 \over (4\pi)^2}\int dk^{2}\,\ln (1 + {(\bar{\Phi}\Phi)-
\sqrt{((\bar{\Phi}\Phi)^2 -\Phi^2\bar{\Phi}^2)\mathstrut}\over
4k^2})=\nonumber\\
&=&-{1\over (8\pi)^{2}}(\Phi\bar{\Phi})\left(
\ln {\Phi^{2}\bar{\Phi}^{2}\over 16{\rm e}^{2}\Lambda^{4}}+
s\ln {1+s\over 1-s}\right)\label{khyp},
\eea
where we have used the notation
$
s^2= 1-{\Phi^{2}\bar{\Phi}^{2}\over (\Phi\bar{\Phi})^{2}} < 0.
$\\
2) The effective action $\Gamma_{\rm SYM}=\Gamma_{\rm V}+ \Gamma_{\rm GD}$
induced by ${\cal N}=2$ vector multiplet contains vector loop contribution
\bea\label{kvec}
&K_{\rm V}=-{\displaystyle \int {dk^{2}\over (4\pi)^2}\,\left( \ln(1+{(\Phi\bar{\Phi})\over k^2})+
\ln (1+{(\Phi\bar{\Phi})-\sqrt{\Phi^{2}\bar{\Phi}^2} \over 2k^2}) +
\ln (1+ {(\Phi\bar{\Phi})+\sqrt{\Phi^{2}\bar{\Phi}^2} \over 2k^2}) \right)}=&
\nonumber\\
&={\displaystyle {1\over (4\pi)^2}}\left((\Phi\bar{\Phi})\ln {\displaystyle {\Phi^{2}\bar{\Phi}^{2}\over {\rm e}^{2}\Lambda^{4}}}+
(\Phi\bar{\Phi})\ln t + \sqrt{\,\Phi^{2}\bar{\Phi}^{2}}\left[{\displaystyle {t+1\over 2}\ln {t+1\over 2} +
{t-1\over 2}\ln {t-1\over 2}}  \right] \right),\label{vector_term}&
\eea
where the notation $t={{\Phi}\bar{\Phi}\over \sqrt{{\Phi}^{2}\bar{\Phi}^{2}}}$
was introduced, plus\\
3) The gauge dependent contribution
\bea\label{kgaug}
K_{\rm GD}&=&\int {dk^2\over (4\pi)^2} \ln \left(1+ {((\bar{\Phi}\Phi)^2
-\Phi^2\bar{\Phi}^2)\over (k^2+\lambda
(\bar{\Phi}\Phi))(k^2+\bar{\lambda}(\bar{\Phi}\Phi))}
\left[-{\lambda\bar{\lambda}\over 2}+\alpha \left({\lambda\bar{\lambda}\over
4k^2 }(\bar{\Phi}\Phi)+ {\lambda+\bar{\lambda}\over
2}-{\alpha\over 4 } \right) \right]\right)=\nonumber\\
&=&\int {dk^2\over (4\pi)^2} \ln \left({
(k^{2} - e_{1})(k^2 - e_{2})(k^2 - e_{3})\over k^{2}(k^{2}+1)^2}
\right),
\eea
which automatically vanishes for abelian background fields $\Phi$. This
is the main result of the subsection. Dependence of the one-loop
effective action on all gauge parameters is given by this expression.

When $\lambda=\bar{\lambda}=0$ the result (\ref{khyp}, \ref{kvec}, \ref{kgaug})
coincides with one given in ref. \cite{grro}. The case $\lambda=0,\,\alpha=1$
is known as Fermi gauge. The corresponding form of K\"ahlerian potential
(\ref{khyp}, \ref{kvec}, \ref{kgaug}) was found in ref. \cite{piwe}. Result for
Landau-DeWitt gauge is obtained at $\alpha=0,\, \lambda=\bar{\lambda}=1$.
Note that (\ref{kgaug}) in the gauge $\alpha=\lambda=\bar{\lambda}=1$,
which can be naturally called as Fermi-DeWitt, two last
terms in the first line (\ref{kvec}) are exactly cancelled by (\ref{kgaug})
while the first term (\ref{kvec}) is being doubled.

Using the integral (\ref{integ}) in (\ref{kgaug}) we obtain
\be\label{rorep}
K_{\rm GD}=e_{1}\ln (-e_{1})+ e_{2}\ln (-e_{2})+ e_{2}\ln (-e_{2}),
\ee
where $e$'s are the roots of the polynomial, which appears in process
of integration of (\ref{kgaug})
\be\label{roots} e_{1}=-{2\over
3}+{1\over 6}R_{1},\quad e_{2}=-{2\over 3}-{1\over
12}R_{1}+{i\sqrt{3}\over 12}R_{2}, \quad e_{3}=e_{2}^{*}.
\ee
Finding the roots of the polynomial and using eq. (\ref{rorep}) we get
the final result for (\ref{kgaug}):
\be\label{kgd}
K_{\rm GD}(s^2,\gamma)=\Phi\bar{\Phi}\tilde{K}_{\rm GD}=\lambda
{\Phi\bar\Phi\over (4\pi)^{2}}\left(-{2\over 3}\ln (\gamma
{s^2\over 4}) + {R_{1}\over 12}\ln ({e_{1}^{2}\over
|e_{2}|^{2}})+{i\sqrt{3}R_{2}\over 12}\ln ({e_{2}\over
e_{2}^{*}})\right),
\ee
\bea
&R_{1}=\Delta_{1}+ \Delta_{2} \quad R_{2}=\Delta_{1} - \Delta_{2},\quad
\Delta_{1,2}=\left(-b \pm \sqrt{-a^{3}+b^2}\right)^{1\over 3},& \nonumber\\
& a=4+3s^{2}(\gamma^{2}-4\gamma+2),
\quad b=-8+9s^{2}(2\gamma^{2}-5\gamma+4),\quad \gamma={\alpha\over \lambda},
&\nonumber
\eea
and we have assumed $\lambda=\bar{\lambda}$. This
form of gauge-dependent part of K\"ahler potential allows to investigate a
gauge dependence of non-holomorphic effective potential ${\cal H}$.

\subsection{${\cal N}=2$ non-holomorphic potential}
In previous subsection we have found the one-loop K\"ahler effective
potential $K(\Phi,\bar{\Phi})$ induced by both ${\cal N}=2$ vector
multiplets and hypermultiplets. As it has been mentioned in refs.
\cite{wigr, piwe}, the K\"ahler potential in the nonabelian case determines
not only by the holomorphic function ${\cal F}$. The additional terms originate
from a real function ${\cal H}({\cal W}, \bar{\cal W})$ of
the ${\cal N}=2$ Yang-Mills superfield strength ${\cal W}$, which is
integrated over full ${\cal N}=2$ superspace. We can derive the
one-loop contribution to ${\cal F}$ and ${\cal H}$ comparing the
last term in decomposition (\ref{full}) with K\"ahler potential. It
leads to
\be\label{recov}
K(\Phi,\bar{\Phi})=\bar{\Phi}^{A}{\cal
F}_{A}+\Phi^{2}(\bar{\Phi}^{A}{\cal H}_{A}) -
(\Phi\bar\Phi)(\Phi^{A}{\cal H}_{A}).
\ee
where
$$
\Phi^{A}{\cal H}_{A}=0,\quad \bar{\Phi}^{A}{\cal H}_{A}=
-{2\bar{\Phi}^2\over (\Phi\bar{\Phi})}s^{2}{\partial {\cal H}\over\partial s^{2}}.
$$

It is well known that $\beta$-function and axial anomaly exactly arise from
holomorphic potential ${\cal F}$. This fact gives us a unique recept
for extracting contributions from K\"ahler potential, which can be associated
with holomorphic and non-holomorphic potentials respectively.

Using the expressions (\ref{khyp}) and (\ref{kvec}) and the
reconstruction formula (\ref{recov}), ones find, in accordance
with ref. \cite{wigr}, the contributions to holomorphic potential
${\cal F}({\cal W})$ and to non-holomorphic potential ${\cal
H}({\cal W}, \bar{\cal W})$ depending on the ${\cal N}=2$
superfield strength ${\cal W}$: \be\label{fhyp} {\cal F}_{\rm
Hyper}^{fund} = {-1\over (8\pi)^2}{\cal W}^2\ln {{\cal W}^2 \over
{\rm e}^2 \Lambda^2} \ee \be\label{fvec} {\cal F}_{\rm Vector} =
{1\over (4\pi)^2}{\cal W}^2\ln {{\cal W}^2 \over {\rm e}^2
\Lambda^2} \ee \be\label{hhhyp} {\cal H}_{\rm Hyper}= {1\over
(16\pi)^2} \ln^2 {1+s \over 1-s}, \ee \be\label{hvec} {\cal
H}_{\rm Vector}= {-1\over (8\pi)^2}\left( \int^{t^2}_{0}du\,{\ln
u\over u-1} + 2\ln {t+1\over 2}\ln {t-1 \over 2}\right). \ee It is
interesting to point out that eq. (\ref{hvec}) can be exactly
rewritten in terms of Euler dilogarithm function ${\rm Li}_{2}(t)$
(see, e.g. ref.  \cite{batem}) \be {\cal H}_{\rm Vector} =
{-1\over (8\pi)^2}\left( - {\rm Li}_{2}(1-t^{2}) + 2\ln {t+1\over
2}\ln {t-1 \over 2}\right). \ee

Our further aim is to obtain off-shell gauge-dependent contribution to ${\cal H}$
from the gauge-dependent part of the full K\"ahler potential. In this
case eq. (\ref{recov}) is written in the form
\be\label{hdep}
-2s^{2}(1-s^{2}){d{\cal H}_{\rm GD}\over ds^{2}}= \tilde{K}_{\rm GD}(s^{2}),
\ee
where $\tilde{K}_{\rm GD}$ was introduced in eq. (\ref{kgd}),
$s^2=1-1/t^{2}$ and
$t={{\cal W}\bar{\cal W}\over \sqrt{{\cal W}^{2}\bar{\cal
W}^{2}}}$, $t\in [0,1]$.  It has already been noticed that $K_{\rm GD}
=0$ at $s^{2}\rightarrow 0$ and therefore ${\cal H}_{\rm GD}$
vanishes on-shell.

We see the holomorphic potential ${\cal F}$ is gauge independent.
All dependence on gauge-fixing parameters is concentrated in the
term ${\cal H}_{\rm GD}$ of non-holomorphic potential ${\cal H}$.
Our next aim is to obtain the ${\cal H}$. The form (\ref{kgd})
of K\"ahler potential is not very convenient for this aim and
further analysis because of its complicated structure, though it
reproduces all known results as partial cases. Therefore we
reformulate (\ref{kgd}) to more simple and suitable form using the
special algebraic methods. Let us present (\ref{kgd}) as a formal
power series.  Eq. (\ref{rorep}) is nothing but a determination of
a symmetrical function via the polynomial roots. According to the
fundamental theorem in theory of symmetrical functions (see e.g.
ref.  \cite{vdwaer}) "any entire rational symmetrical function can
be uniquely rewritten as a entire rational function of elementary
symmetrical functions" (i.e. coefficients of the polynomial).

To represent (\ref{rorep}) as an entire rational function we expand the logarithms
into a formal power series
\be\label{gdsum} K_{\rm GD} = -\sum_{n=1}^{\infty}{1\over n}\,S_{n},
\ee
where the {\it power} symmetrical functions of the roots $e_{1}, e_{2}, e_{3}$
of the form
\be\label{defsn}
S_{n} = e_{1}(1 + e_{1})^{n} + e_{2}(1 + e_{2})^{n} + e_{3}(1 + e_{3})^{n}
\ee
has been used. Using classical recursion Newton's formulae we can uniquely
express $S_{n}$ in terms of elementary symmetrical functions.

It is well known that the roots $e_{i}$ of an algebraic equation are always
satisfy the Vieta relations.
For the roots (\ref{roots}) of the polynomial, which appears from the
numerator in the logarithm of (\ref{kgaug}) we have
\be\label{vieta}
-e_{1}e_{2}e_{3}=g_{3},\quad
e_{1}e_{2} + e_{2}e_{3} + e_{1}e_{3} = g_{2}, \quad
e_{1} + e_{2} + e_{3} = -2,
\ee
where elementary symmetrical functions are given from (\ref{roots}, \ref{vieta})
as $ g_{2}=1+s^2(-{1\over 2}+\gamma(1-{\gamma\over 4})),\quad
g_{3}=s^2{\gamma\over 4}.$

Multiplying eq. (\ref{defsn}) by $e_{1} + e_{2} + e_{3}$ and using
identities (\ref{vieta}) we obtain the recursion relation
\be
S_{n+1} - S_{n} - (1 - g_{2})S_{n-1}+(1-g_{2}+g_{3})S_{n-2} = 0.
\ee
Using this relation one can evaluate any $S_{n}$ step by step.
Writing out the few first symmetrical functions
\bea\label{first_terms}
&S_{1}=2(1-g_{2}),\,S_{2}= -2(1-g_{2}) - 3g_{3},\, S_{3}= - 2(1-g_{2})^{2} - g_{3},\,& \nonumber\\
&S_{4}=-6(1-g_{2})^{2} - 6(1-g_{2})g_{3} - g_{3},\ldots&
\eea
one can see that $S_{n}\sim s^{2}$ for any $n$. It allows to  simplify
integration in eq. (\ref{hdep}). In addition, we note that each $S_{n}$
includes $g_{3}$ linearly.

Moreover, the known Waring formulae (see. e.g. \cite{vdwaer})
allow to express $S_{n}$ for any $n$ directly in terms of
$g_{2},\,g_{3}$. In order to get all $S_{n}$, it is very useful to
introduce a generating function defined by a formal power series
\be\label{genfunc} G(\tau) =\sum_{k=1}^{\infty}\,\tau^{k-1}S_{k},
\ee then any $S_{n}$ can be found with help of differentiations of
the generating function $G$ with respect to $\tau$. It also allows
us to express a general term of the sequence $S_{n}$ in terms of
symmetrical functions $g_{2}$ and $g_{3}$ instead of the roots
$e_{i}$. Since the functions $g_{2},\,g_{3}$ are known from the
integral (\ref{kgaug}), we can avoid finding the roots $e_{i}$ for
analysis ${\cal H}_{\rm GD}$ at all.

The generating function $G$ satisfies an algebraic equation
which can be derived by multiplying the recursion relation by
$\tau^{k}$ and summing over powers $k$.
The solution to this equation is
\be\label{genf}
G(\tau) = {2(1 - g_{2}) - 4(1 - g_{2})\tau -
3g_{3}\tau +2(1 - g_{2}+ g_{3})\tau^{2} \over 1 - \tau - (1 -
g_{2})\tau^{2} + (1 - g_{2} + g_{3})\tau^{3}}.
\ee
As a result we obtain an expansion of $K_{\rm GD}$ in terms of
elementary symmetrical functions $g_{2}$, $g_{3}$:
\be\label{defrep}
K_{\rm GD} = -\sum_{n=1}^{\infty}{1\over n!} \left({d\over d\tau}\right)^{n-1}
G(\tau)|_{\tau=0}.
\ee
Now, it is useful to introduce the new parameters
$g=-1/2+(\gamma/2-1)^{2},\, g_{3}=\gamma/4,\,p=g+g_{3}$, $u=1-s^{2}$.
Using the binominal formula for derivatives of the generating function
(\ref{genf}) in (\ref{defrep}), we rewrite the equation (\ref{hdep}) in
the following form
\be\label{hyrow}
-2u{d{\cal H}_{\rm GD}\over du}=\sum_{k=0}^{\infty}{1\over (k+3)!}
\left(4g-g_{3}(k+1)(k+5)\right)\left({d\over d\tau}\right)^{k}Y|_{\tau=0},
\ee
where
$
Y^{-1}=1-\tau-g\tau^2+p\tau^3+u(g\tau^2-p\tau^3).
$
It is useful to extract, in the right hand side of eq. (\ref{hyrow}),
the powers of $u$ and rewrite this relation in form of double sum
\be\label{hrow}
-2u{d{\cal H}_{\rm GD}\over
du}=\sum_{l=0}^{\infty}u^{l}\sum_{k=0}^{\infty}{k!\over (k+3)!}
\left(4g-g_{3}(k+1)(k+5)\right)\left[{1\over k!}\left({d\over
d\tau}\right)^{k}Q_{l}\right]_{\tau=0},
\ee
and
$$
Q_{l}={(-g\tau^2+p\tau^3)^{l}\over (1-\tau-g\tau^2+p\tau^3)^{l+1}}.
$$
Last expression allows to find ${\cal H}_{\rm GD}$ as a series with a
coefficient, at each given power of $u$, depending on elementary
symmetrical functions. Hence, we finally can rewrite (\ref{hdep})
in terms of elementary symmetrical functions. We see that the right
hand side (\ref{hrow}) can be written via rational functions for any
given choice of gauge parameters. For some partial choice of gauge
parameters, arbitrary term of series can be found exactly. Therefore,
directly finding several first derivations
$\left({d\over d\tau}\right)^{k}Q_{l}$ at $\tau=0$ (for example by
{\it Mathematica} software) in right side (\ref{hrow}) one can
restore the general term of the series (\ref{hrow}) and then
directly fulfil summation over powers $u$. As a result, the
gauge-dependent part of effective action can always be written in
{\it any given gauge} as a series over powers of elementary
symmetrical functions. The actual summation of such a series can
be realized for any specific choices of gauge parameters.
We point out that the procedure described above can be used, in
principle, for evaluating a functional determinant for an arbitrary
higher order non-minimal operator in the low-energy approximation.
Actually, the only we need is a system of roots of a polinomial
corresponding to the operator in the momentum representation.

For example, taking only linear in $\alpha$ and $\lambda$ terms in
(\ref{hrow}) we obtain the few first terms in the expansion of
${\cal H}_{\rm GD}$ for arbitrary $\alpha,\,\lambda$:
\be\label{hasimp}
{\cal H}_{\rm GD}(\lambda, \alpha)={1\over (4\pi)^{2}}\left[\lambda \left(\sqrt{2}\ln (1+\sqrt{2})-\ln(2)\right)+
{\alpha\over 2}(3-\ln (\varepsilon)) + \cdots\right]\ln (t) + O(t),
\ee
where $\varepsilon$ is IR cut-off parameter. The occurrence of IR-divergence
could be seen directly from (\ref{gdsum}) and (\ref{first_terms}). As it has
been mentioned above $S_{n}\sim g_{3}$ and therefore the corresponding series
is divergent.  But, for $\alpha=0$ the IR-divergence does not appear
(i.e. in the Landau-DeWitt gauge: $\alpha=0, \, \lambda=1$).

Let's consider the Landau-DeWitt gauge in more detail. At such a
choice $Y^{(k)}$ in (\ref{hyrow}) becomes enough simple
\be
Y^{(k)}=k!\left({1\over 1-a^{2}}-{(-a)^{k+1}\over
2(1+a)}-{a^{k+1}\over 2(1-a)} \right), \quad a^{2}={s^{2}\over 2}
\ee
and the general term in right side (\ref{hyrow}) can be exactly
found. For example
\be
\sum_{k=0}^{\infty}{k!\,a^{k}\over (k+3)!}=
{1\over 6}\,{}_{2}F_{1}(1,1;4;a)= {1\over
4a^{3}}\left[a(3a-2)-2(1-a)^{2}\ln (1-a)\right], \ee and \be
\sum_{k=0}^{\infty}{k!\over (k+3)!}= {1\over
6}\,{}_{2}F_{1}(1,1;4;1)= {1\over 4},
\ee
where ${}_{2}F_{1}$ is the Gauss hypergeometric function (see, e.g. \cite{batem}).

Finally, the expression (\ref{hyrow}) becomes
\be\label{case0}
(1-2a^{2}){d{\cal H}_{\rm GD}\over da}={1-a\over a}\ln (1-a)+{1+a\over a}\ln (1+a)
\ee
and we obtain ${\cal H}_{\rm GD}$ by integration
\bea
2(4\pi)^{2}{\cal H}_{\rm GD}&=&
\ln (2)\ln (1-s^{2})+{1\over \sqrt{2}}\ln \left({\sqrt{2}-1\over \sqrt{2}+1}\right)
\ln (1-s^{2})-{\rm Li}_{2}\left({s^{2}\over 2}\right)+\nonumber\\
&+&{\sqrt{2}-1\over\sqrt{2}}
\left[{\rm Li}_{2}\left({s-1\over\sqrt{2}-1}\right)+
{\rm Li}_{2}\left(-{s+1\over\sqrt{2}-1}\right)\right]+\nonumber\\
&+&{\sqrt{2}+1\over\sqrt{2}}
\left[{\rm Li}_{2}\left({s+1\over\sqrt{2}+1}\right)+
{\rm Li}_{2}\left({1-s\over\sqrt{2}+1}\right)\right]\label{gdfin1}
\eea
It should be noted that for the considered gauge, ${\cal H}_{\rm GD}$ can
be also found by direct integration from (\ref{kgaug}) with the same
result (\ref{gdfin1}). That proves correctness of the described above method.
We remind that
$
s^2= 1-{{\cal W}^{2}\bar{\cal W}^{2}\over ({\cal W}\bar{\cal W})^{2}} \le 0.
$
Of course, on-shell, where ${\cal W}$ and $\bar{\cal W}$ are abelian,
we have $s=0$ and therefore ${\cal H}_{\rm GD}$ vanishes as the other
contributions to ${\cal H}$.

We emphasize that expressions (\ref{hhhyp}, \ref{hvec}) and (\ref{gdfin1}) are exact
results within one-loop approximation. Of course, they can
be expanded in serieses in two limit cases: $t\rightarrow 1$ and
$t\rightarrow 0$. At $t\rightarrow 1$ (almost abelian case) we have
\bea
{\cal H}_{\rm Vector}&\sim&{-1\over (8\pi)^2}\left( (t-1)\ln ({t-1\over
2})+\ldots\right),\nonumber\\ {\cal H}_{\rm Hyper}&\sim&{1\over
(8\pi)^2}\left(2(t-1) -{1\over 3}(t-1)^2 +\ldots\right).
\eea
To analyze the analytical properties of ${\cal H}_{\rm GD}$ near of
this point it is useful to present solution to eq. (\ref{case0}) by the
series
\be\label{alphsum}
2(4\pi)^2{\cal H}_{\rm GD} = -{1\over 2}\ln
(1-s^{2})+ \sum_{k=1}^{\infty}\, \left({s^{2}\over 2}\right)^{k+1}
{{}_{2}F_{1}(1,k+1;k+2;s^{2})\over (2k+1)(k+1)^{2}}.
\ee
This representation is defined when $|s^2|<1$ and provides the obvious power
expansion.  Moreover, we can analytical extend ${\cal H}_{\rm GD}$ in
the complex plane with the cut $[1,\,\infty]$ in order to analyze its
behavior near of all branching points. For
example, from the known identity
$$
{}_2F_{1}(1,k+1;k+2;s^2)=(k+1){\bf \Phi}(s^2,1,k+1),
$$
where ${\bf \Phi}$ is so called the Lerch transcendental function
(see, e.g. \cite{batem}). Taking into account that
$lim_{s^2\rightarrow 1}{\bf \Phi}(s^2,1,k+1)/ln(1-s^2)=-1$ we get
at $s^2 \rightarrow 1$ (or $t^2\rightarrow -\infty$) the
logarithmical branch point \be {\cal H}_{\rm GD}\sim \ln
(1-s^2)(ln (2)+{1\over \sqrt{2}} \ln ({\sqrt{2}-1\over
\sqrt{2}+1}))+\ldots \ee At small $t$, large $-s^{2}$
(equivalently, large "mass") we have an asymptotic
Schwinger-DeWitt perturbation series in inverse powers of "mass"
$\sqrt{{\cal W}^{2}\bar{\cal W}^{2}}$ and a logarithmic branch
point, that can be seen from another representation of
eq.(\ref{alphsum}) \be 2(4\pi)^2{\cal H}_{\rm GD} = -{1\over 2}\ln
(t^{2})+ \sum_{k=1}^{\infty}\, \left({t^{2}-1\over 2}\right)^{k+1}
{{}_{2}F_{1}(k+1,k+1;k+2;1-t^{2})\over (2k+1)(k+1)^{2}}. \ee
Summation of the asymptotic series at $t\sim 0$ leads to \bea
{\cal H}_{\rm GD}& \sim -{1\over (8\pi)^2}(t^2\ln (t^2) + \sqrt{2}
\pi t +\ldots) \eea In this region on the isolated branch of the
multifunction we also have \bea {\cal H}_{\rm
Vector}&\sim&{-1\over
(8\pi)^2}\left(2i\pi t -t^2 \ln (t^2)+\ldots\right),\nonumber\\
{\cal H}_{\rm Hyper}&\sim&{1\over (8\pi)^2}\left(\pi t -t^2
+\ldots\right). \eea Let us remember that the appearance of the
imaginary part related to the second term in eq. (\ref{hvec}),
which (as it has been mentioned earlier) is missing in the
Fermi-DeWitt gauge. Such a behavior is not unusual and it looks
like quite analogous to a well-known exactly solvable model in
effective field theory --- namely the Euler-Heisenberg effective
action. It is pointing out some more property of the
Euler-Heisenberg effective action at small mass (strong external
field), it possesses by logarithmic branch point as well as ${\cal
H}_{\rm GD}$, ${\cal H}_{\rm Vector}$ while at large mass (weak
external field) there exists an asymptotic series expansion in
inverse powers of mass (see e.g.  \cite{land}).

We showed that the gauge-dependent part of off-shell effective
action can be found with an arbitrary level of accuracy and at any
choice of gauge fixing parameters. The form of the non-holomorphic
effective potential has an essential arbitrariness due to its
explicit gauge dependence. In particular, this fact leads to the
ambiguous definition of
$R_{A\bar{B}C\bar{D}}(W^{A\alpha}W^{C}_{\alpha}\bar{W}^{B\dot{\alpha}}
\bar{W}^{D}_{\dot{\alpha}})$ term from eq. (\ref{full}), which
should reproduce the leading term in the expansion of the
non-abelian analog of the Born-Infeld action (see, e.g.
\cite{ts}). The structure of the tensor $R_{A\bar{B}C\bar{D}}$ is
enough cumbersome. Besides this we point out that the symmetrized
trace $(F^{+})^{2}(F^{-})^2/\phi^{2}\bar{\phi}^{2}$, defying the
full set of $F^{4}$-terms in effective action also contains the
various contractions $\phi^{A},\,\bar{\phi}^{A}$ with $F^{A}$.
Existence a large class of gauge theory operators, which
correspond to supergravity modes and contain nontrivial extra
factors (depending on $\phi^{A},\,\bar{\phi}^{A}$), in non-abelian
Born-Infeld action was discussed in refs. \cite{ts, fer}.

To conclude this subsection we note that unlike the abelian
case, ${\cal N}=2$-supersymmetry itself can not uniquely fix a
form of next-to-leading term in the effective action because of its
explicit gauge dependence.

\section{Summary}
We have studied the non-holomorphic potential depending on non-abelian
strengths ${\cal W}$ and $\bar{\cal W}$ in ${\cal N}=2$ supersymmetric
theory of $SU(2)$ gauge multiplet coupled to hypermultiplet. The theory
under consideration was realized in terms of ${\cal N}=1$ superfields
and described by ${\cal N}=1$ gauge superfield interacting with three
chiral superfields in some representations of gauge group. We quantized
the theory withing background field method using three-parametric
${\cal N}=1$ supersymmetric $R_{\xi}$-type gauge and constructed the
corresponding quadratic action describing the one-loop effective
action.

We have calculated the K\"{a}hlerian effective potential depending on
${\cal N}=1$ chiral superfield projection of ${\cal N}=2$ superfield
strength taking the values in Lie algebra of $SU(2)$ group and
containing all three gauge parameters. Using the special methods of
the polinomial algebra we developed a general recurrent procedure of
obtaining manifestly ${\cal N}=2$ supersymmetric non-holomorphic
potential for a class of gauge parameters under consideration. This
potential reproduces all previous results on non-holomorphic potential
in non-abelian background as partial case. The procedure we have
developed to compute the gauge dependent contribution to effective
one-loop effective action is quite generic and, in principle, it can be
considered as a new method of calculating an one-loop effective action
for arbitrary (non-minimal, higher order) differential operators.

The special case of supersymmetric Landau-DeWitt gauge was
investigated in more details. It is turned out to be that the
non-holomorphic potential is exactly found for this case in terms
of Euler dilogarithm function. We have also studied the various
limiting situations of the non-holomorphic potential, in particular "near
on-shell" limit and "large mass" $\sqrt{\bar{\cal W}^{2}{\cal
W}^{2}}$ limit. It is interesting to point out that Euler
dilogarithms occur in many problems associated with quantum ${\cal
N}=2$ supersymmetric field models (see e.g. ref. \cite{eden})

\section{Acknowledgments}
The authors are grateful to S.J. Gates for the useful comments.
The work was supported in part by INTAS grant, INTAS-00-00254.
I.L.Buchbinder is also grateful to RFBR grant, project No
02-02-04002 and to DFG grant, project No 436 RUS 113/669 for partial
support.  The work of N.G.Pletnev and A.T.Banin was supported in
part by RFBR grant, project No 00-02-17884.

\newpage
\section*{Appendixes}
\appendix
\setcounter{equation}{0}
\renewcommand{\theequation}{\thesection. \arabic{equation}}

\section{ ${\cal{N}} =1$ structure of ${\cal N}=2$ superfield strengths}
The ${\cal N}=2$ SYM theory is usually formulated in the ordinary ${\cal N}=2$
superspace by imposing certain constrains on the gauge and super
covariant derivative $\nabla_{\alpha a}$ and $\bar{\nabla}_{\dot{\alpha}}^{a}$
\be
\{\nabla_{\alpha a},\nabla_{\beta b}\}=iC_{ab}C_{\alpha\beta}\bar{\cal W}, \quad
\{\bar{\nabla}_{\dot{\alpha}}^{a},\bar{\nabla}_{\dot{\beta}}^{b}\}=
iC^{ab}C_{\dot{\alpha}\dot{\beta}}{\cal W},\quad
\{\nabla_{\alpha a},\bar{\nabla}_{\dot{\beta}}^{b}\}=i\delta_{a}^{b}\nabla_{\alpha\dot{\beta}},
\ee
where ${\cal W}, \bar{\cal W}$ are chiral (antichiral) scalar superfield
strengths, respectively (see ref. \cite{ggrs}). The ${\cal N}=2$ chiral superfield
$ \bar{\nabla}_{\dot{\alpha}a}{\cal W}=0$
is reducible, unlike its ${\cal N}=1$ counterpart. To achieve an irreducible superfield we may additionally
impose important the constrains of reality condition
\be
\nabla^{\alpha}_{a}\nabla_{\alpha b}{\cal W} =
C_{ac}C_{bd}\bar{\nabla}^{\dot{\alpha} d}\bar{\nabla}^{c}_{\dot{\alpha}}\bar{\cal W}.
\ee
The rigid $SU(2)_{R}$ indices are raised or lowered with the help of
the antisymmetric invariant tensor $C_{ab}$, with $C_{12}=C^{12}=1$.
It is customary to represent the solution to these constraints in the
form of an ${\cal N}=1$ superfield expansion. We defines ${\cal N}=1$ superfield
components both in the adjoint representation nonabelian gauge group
as ${\cal W}|=\phi$, $\nabla_{\alpha 2}{\cal W}|=-W_{\alpha}$
etc. The bar denotes setting $\theta^{2\alpha}=\eta^{\alpha}=0$,
$\theta^{\dot{\alpha}}_{2}=\bar{\eta}^{\dot{\alpha}}=0$. As a result of
reducing ${\cal N}=2$ superfield to ${\cal N}=1$ form we find:
\bea
{\cal W}^{A}&=&\Phi^{A}-\eta^{\alpha}W_{\alpha}^{A}-\eta^{2}\bar{\nabla}^{2}\bar{\Phi}^{A}+
i\eta^{\alpha}\bar{\eta}^{\dot{\alpha}}\nabla_{\alpha\dot{\alpha}}\Phi^{A}+
\eta^{2}\bar{\eta}^{\dot{\alpha}}(i\nabla^{\alpha}_{\dot{\alpha}}W_{\alpha}^{A}+
f^{ABC}\Phi^{B}\bar{\nabla}_{\dot{\alpha}}\bar{\Phi}^{C})+\nonumber\\
&+&\bar{\eta}^{2}\eta^{\alpha}f^{ABC}\Phi^{B}\nabla_{\alpha}\Phi^{C}+
\eta^{2}\bar{\eta}^{2}(\Box\Phi^{A}+ {1\over
2}f^{ABC}\Phi^{B}\bar{\nabla}^{\dot{\alpha}}\bar{W}_{\dot{\alpha}}^{C}-
f^{ABC}(\nabla^{\alpha}\Phi^{B})W_{\alpha}^{C}-\nonumber\\
&-&f^{ABC}f^{DEC}\Phi^{B}\Phi^{D}\bar{\Phi}^{E})\nonumber\\
\bar{\cal W}^{A}&=&\bar{\Phi}^{A}-\bar{\eta}^{\dot{\alpha}}\bar{W}_{\dot{\alpha}}^{A}-
\bar{\eta}^{2}\nabla^{2}\Phi^{A}-i\eta^{\alpha}\bar{\eta}^{\dot{\alpha}}
\nabla_{\alpha\dot{\alpha}}\bar{\Phi}^{A}+
\bar{\eta}^{2}\eta^{\alpha}(i\nabla_{\alpha}^{\dot{\alpha}}\bar{W}_{\dot{\alpha}}^{A}-
f^{ABC}\bar{\Phi}^{B}\nabla_{\alpha}\Phi^{C})+\nonumber\\
&+&\eta^{2}\bar{\eta}^{\dot{\alpha}}f^{ABC}\bar{\Phi}^{B}
\bar{\nabla}_{\dot{\alpha}}\bar{\Phi}^{C}+
\eta^{2}\bar{\eta}^{2}(\Box\bar{\Phi}^{A}+ {1\over
2}f^{ABC}\bar{\Phi}^{B}\nabla^{\alpha}W_{\alpha}^{C}-
f^{BCA}(\bar{\nabla}^{\dot{\alpha}}\bar{\Phi}^{B})\bar{W}_{\dot{\alpha}}^{C}-\nonumber\\
&-&f^{ABC}f^{CDE}\bar{\Phi}^{B}\bar{\Phi}^{D}\Phi^{E}).\label{expa}
\eea
Component expansion ${\cal N}=1$ superfields is well
known. Namely this eq. (\ref{expa}) has been used to derive ${\cal N}=1$ form of
low-energy effective action (\ref{full}) . Another way of derivating was given
in ref. \cite{wigr}.

\end{document}